\documentclass[11pt, letterpaper]{article}

\usepackage{etoolbox}

\usepackage[letterpaper,margin=1in]{geometry}

\usepackage{url}
\usepackage{microtype}
\usepackage{chngcntr}
\usepackage[plain]{algorithm}
\usepackage{algpseudocode}
\usepackage{fixmath} 
\usepackage{amsmath} 
\usepackage{amsthm}
\usepackage{amssymb} 
\usepackage{times}  
\usepackage{epsfig}
\usepackage{tabularx}
\usepackage{graphicx} 
\usepackage{color}
\usepackage{xspace}
\usepackage{thumbpdf}
\usepackage{listings}
\usepackage{verbatim}
\usepackage{booktabs}
\usepackage{colortbl}
\usepackage{url}
\usepackage{afterpage}
\usepackage{balance}
\usepackage{xcolor}

\usepackage{epigraph}

\usepackage{hyperref}

\makeatletter
\let\MYcaption\@makecaption
\makeatother

\usepackage[font=footnotesize]{subcaption}
\usepackage[font=footnotesize]{caption}
\usepackage{enumitem}

\makeatletter
\let\@makecaption\MYcaption
\makeatother

\newtheorem{theorem}{Theorem}

\newtheorem{observation}{Observation}

\definecolor{Ora}{cmyk}{0, 0.6, 0.8, 0}
\definecolor{carribgreen}{rgb}{0.0, 0.8, 0.6}
\definecolor{mygray}{gray}{0.5}

\newcommand{\hide}[1]{}

\providecommand{\ie}{\emph{i.e.,} }
\providecommand{\eg}{\emph{e.g.,} }

\providecommand{\etal}{\emph{et al. }}

\newenvironment{myabstract}
{\list{}{\listparindent 1.5em
		\itemindent    \listparindent
		\leftmargin    1cm
		\rightmargin   1cm
		\parsep        0pt}
	\item\relax}
{\endlist}

\newenvironment{mycover}
{\list{}{\listparindent 0pt
		\itemindent    \listparindent
		\leftmargin    1cm
		\rightmargin   1cm
		\parsep        0pt}
	\raggedright
	\item\relax}
{\endlist}

\newcommand{\myaff}[1]{\,$\cdot$\, {\small #1}\par\medskip}

\hypersetup{
	colorlinks=true,
	linkcolor=black,
	citecolor=black,
	filecolor=black,
	urlcolor=[rgb]{0,0.1,0.5},
	pdftitle={ Distributed Consistent Network Updates in SDNs},
	pdfauthor={Klaus-Tycho Foerster and Stefan Schmid}
}

\begin{document}

\begin{mycover}
	{\LARGE\bfseries\boldmath Distributed Consistent Network Updates in SDNs: Local Verification for Global Guarantees \par}
	\bigskip
	\bigskip

	\textbf{Klaus-Tycho Foerster}
	\myaff{University of Vienna, Austria}

	\textbf{Stefan Schmid}
	\myaff{University of Vienna, Austria}
\end{mycover}

\medskip
\begin{myabstract}
	\noindent\textbf{Abstract.}
While SDNs enable more flexible and adaptive network operations,
(logically) centralized reconfigurations introduce overheads
and delays, which can limit network reactivity.
This paper initiates the study of a more distributed approach,
in which the consistent network updates are implemented by the switches
and routers directly in the data plane.
In particular, our approach leverages concepts from local proof
labeling systems, which allows the data plane elements to locally
check network properties, and we show that this is sufficient to obtain
global network guarantees. We demonstrate our approach considering three fundamental use cases, and analyze its benefits in terms of performance and fault-tolerance.
\end{myabstract}

\vspace{12cm}
\begin{myabstract}
\noindent \copyright{} 2019 IEEE\@. This is the authors' version of a paper that will appear in the Proceedings of the 18th IEEE International Symposium on Network Computing and Applications (NCA 2019).
\end{myabstract}

\newpage

\section{Introduction}\label{sec:introduction}

Given the increasingly stringent requirements on the
dependability and performance of communication networks,
it becomes important that networks be able to flexibly
adapt to their context,
e.g., react to failures or to changes in the demand,
in an \emph{automated} manner.
Software-Defined Networks (SDNs) provide such flexibilities
by allowing to update network configurations programmatically,
disburdening human operators
from their most complex tasks  and significantly improving reaction times.
Indeed, over the last years, the algorithmic problem of how to
update networks consistently has received much attention~\cite{survey}.

However, while outsourcing and consolidating the control 
over switches and routers provides great flexibilities, 
indirection via (remote) controllers comes with overheads
in terms of communication and computation costs, and can hence
lead to delays. In fact, it is known that updating
routes in a network while providing even simple transient properties
such as loop-freedom,
requires 
many interactions with the SDN controller 
in the worst case~\cite{DBLP:conf/hotnets/MahajanW13,DBLP:conf/networking/ForsterMW16},
unless one resorts to packet header rewriting.
Given that the control plane
can operate orders of magnitude slower than the data plane~\cite{Feigenbaum:2012:BAR:2332432.2332478},
this is problematic.

This paper investigates opportunities to overcome these overheads
and hence further improve network reactivity. To this end, we explore
a more \emph{distributed} approach to updating routes in networks,
reducing interactions with the control plane \emph{without}
sacrificing flexibility and consistency. 
This is challenging, as without a (logically) centralized network view,
switches and routers need to be able to check certain network properties
\emph{locally}.

We propose and investigate 
the use of distributed mechanisms based on local proof labeling 
systems~\cite{DBLP:conf/sigcomm/SchmidS13}, 
to propagate and implement network updates \emph{entirely in the data plane}.
In particular, we present a solution which allows switches and routers to 
check \emph{locally} if a certain network property is
fulfilled and whether a rule update can be safely applied.
Consequently, a controller (or multiple controllers, in case
of distributed SDN control planes) can simply submit
update requests to the network, which are then propagated
and implemented by the data plane autonomously.
To demonstrate our approach, we consider 
two
fundamental properties,
both 
related to connectivity. 
\begin{itemize}
	\item \emph{Blackhole freedom:}
	There is always a matching rule forwarding a packet 
	to the next hop switch or router.
	\item \emph{Loop freedom:}
 The forwarding rules never contain a loop.
\end{itemize} 

We also evaluate the benefits of our approach 
analytically
and investigate potential speed ups and fault-tolerance.

\vspace{0.1cm}
\noindent\textbf{Contributions.}
This paper presents a distributed approach to
operate and consistently update software-defined
networks, by relying on local proof labeling
systems. We show the feasibility and benefits of our approach on 
two
case studies, demonstrating that using our approach, simple
local verification is sufficient to provide global
correctness guarantees. We also show that our approach
can lead to faster and fault-tolerant network updates. 

\vspace{0.1cm}
\noindent\textbf{Overview.}
The remainder of this paper is organized as follows.
After introducing our model and preliminaries in Sections~\ref{sec:model}, 
we present our main approach in
Section \ref{sec:main-idea}. We discuss 
our two case studies in Section~\ref{sec:blackholes} (considering efficient
updates limited to the affected routes) and 
 Section~\ref{sec:loopfreedom} (removing the need for packet tagging),
and then examine potential speed up and fault-tolerance aspects (Section~\ref{sec:faulttolerance}).
After reviewing related work in Section~\ref{sec:relwork},
we conclude in Section~\ref{sec:conclusion}.

\section{Model and Preliminaries}\label{sec:model}

We  follow standard assumptions~\cite{DBLP:conf/sigcomm/SchmidS13,DBLP:journals/tcs/FoersterLSW18, segway} in our work, both regarding the network and the local verification model.
\vspace{0.04cm}
\noindent\textbf{Network model.} The considered networks are modeled as connected graphs $G=(V,E)$ with $n$  nodes (switches, routers) with unique identifiers and $m$ full-duplex links. 

The network is equipped with a logically centralized controller that can collect the network state and send out conditional network updates to the nodes, \eg changing a forwarding rule once a certain local condition is met~\cite{segway}. 
\vspace{0.1cm}
\noindent\textbf{Local Verification.}
We will also leverage a connection~\cite{DBLP:conf/sigcomm/SchmidS13} between proof labeling schemes~\cite{DBLP:journals/dc/KormanKP10,DBLP:journals/toc/GoosS16} and 
the SDN model~\cite{survey}, see Section~\ref{sec:main-idea}.
A proof labeling scheme can be characterized by a prover-verifier-pair $(\mathcal P,\mathcal V)$ as follows:
Given some property $\mathcal S$ that the network state could uphold after updates (\eg loop freedom), the prover $\mathcal P$ sends new labels to the nodes.
The verifier $\mathcal V$ is a distributed algorithm, running on each node $v$,
that can collect the labels from all neighbors $\mathcal N(v)$.
It outputs \textsc{yes} if property $S$ holds and the labels are from $\mathcal P$, but at least one node must output \textsc{no}, if the property $S$ is violated.

\section{Approach and Main Idea}\label{sec:main-idea}

This section presents 
how to leverage proof labeling schemes in the context of consistent updates for SDNs, both from a methodological and an implementation point of view.

\vspace{0.04cm}
\noindent\textbf{Methodology.}
Many consistency properties are inherently global, \eg long loops cannot be detected by considering the  forwarding rules in the local neighborhood.
Even locally detectable problems can have an impact on nodes far away, such as, \eg a blackhole downstream from the packet source.
We thus utilize the power of proof labeling schemes to allow for \emph{local} verification of consistency properties, also supporting \emph{distributed} consistent network updates.
In our approach,  the controller acts as the prover $\mathcal P$.
Nodes which are aware of the current label state of their neighbors, 
can now check them in the time intervals 
deemed necessary.
In the simplest case, a node informs all its neighbors once its label state changes.
Once being informed about label state changes, nodes can run the verifier $\mathcal V$ to check if the (global) property $S$ is still correct, respectively ring an alarm (\eg to the controller) if not.
The main idea of our approach is that a node will not immediately apply a new label received from the controller, but rather first check if the property $S$ still holds from its point of view after applying said label to itself.
As such, we do not need the large overhead of constantly communicating with the centralized controller regarding the updated network state, but can decide completely locally when to update.
The challenge we undertake in this paper is to actually develop approaches that fulfill these criteria for common consistency properties, \ie generating distributed consistent network updates that can be verified locally.

\vspace{0.1cm}
\noindent\textbf{Implementation.}
Our approach is timely and can be implemented in
OpenFlow and P4-based programmable networks.  
The implementation of the controller is simple
as it only pre-computes the information
needed by the switches later, during the network update
(reducing communication and computation overheads).
Furthermore, our approach does not rely on tight clock
synchronization protocols while providing the 
same benefits~\cite{mizrahi2015timed}. 
In the dataplane, we can use the approach by ez-Segway~\cite{segway},
leveraging per-switch local controllers to manipulate
dataplane state (via OpenFlow).

\section{Efficient Certification Limited to Involved Routes}\label{sec:blackholes}

We first present a solution for efficient certification 
which only involves the nodes along routes that are actually updated
(rather than \emph{all} nodes in the network).

We start with a case study on the \emph{blackhole freedom} property.
A so-called blackhole occurs when a node has no matching rule 
for a packet, \ie the packet is dropped (into a blackhole).
A simple scheme to avoid blackholes for a specific network flow
is to ensure that new labels for a node $v$ always contain a matching rule for the flow destination $d$, 
where an update is rejected otherwise.
However, whereas this scheme is easy to verify and apply, it suffers from the downside that every node in the network must have a forwarding rule for said flow, even if its packets only traverse a small subset of the nodes.
A more efficient solution would supply forwarding rules only to those nodes actually en route, as performed \eg in~\cite{segway} for network flows.
The authors propose a distributed version of the 2-phase update scheme by \eg Reitblatt \etal ~\cite{DBLP:conf/sigcomm/ReitblattFRSW12}\footnote{The 2-phase commit scheme in~\cite{DBLP:conf/sigcomm/ReitblattFRSW12} updates the forwarding for a flow $F$ to $F'$ as follows: 
The new flow rules for $F'$ are distributed in the network, and once ack'ed to the controller, the controller informs the packet source to from now on tag all flow packets with $F'$, instead of the previous tag of $F$.}:
the routing path for flow $F$ is updated in reverse, where the destination informs its predecessor on the path to update its rules for $F'$, which in turn informs its predecessor, and so on.
Eventually, the packet source will be reached, which then knows it is safe to send packets out tagged with $F'$.

Providing verifiable blackhole freedom can be directly achieved in this setting if every node $v$ with a new rule for $F'$ only updates if its successor $w$ on the path has been updated.
Notwithstanding, what cannot be verified so far is the problem of reachability, \ie will the packets in $F'$ actually reach their target?
In the prover-verifier framework, if each node is informed about its successor, a node $w$ could be successor of two nodes $u,v$, which in turn can lead to a forwarding loop.
We can resolve this problem with a construction borrowed from reachability in the context of proof labeling schemes~\cite{DBLP:journals/toc/GoosS16}, by specifying both predecessors and successors of all nodes (besides source and destination). 
Then, by a connectivity argument, the packets of $F'$ cannot loop and will reach the destination when starting from the source.
While we now have verifiable blackhole freedom for the nodes en route, we cannot use the above scheme to actually deploy a new path for $F'$.
Assume that the path has at least two nodes besides the source and the destination, then no further node en route can actually deploy the rules for $F'$ under common asynchrony~\cite{survey} assumptions---both a successor and predecessor along the route is needed.
Moreover, from a structural point of view, such a predecessor-successor construction does not remove unnecessary forwarding loops in the network, \eg a loop disconnected from source/destination cannot be locally detected.
While such disconnected loops might not seem as harmful from a routing point of view, they can hinder future updates and also highlight another downside of the above scheme, namely that it is not suitable for purely destination-based schemes, where routing is performed along a forwarding tree.
We investigate such scenarios in the next section, but first show how to fix our proposed scheme.

To this end, we replace the predecessor-successor relationship with a distance labeling scheme, as described in, \eg \cite{DBLP:journals/dc/KormanKP10,DBLP:journals/toc/GoosS16}.
Each node along the path of $F'$ also obtains its distance to the destination as part of the label, measured in hops along $F'$.
Then, a node will only update if its successor has already updated and its distance is exactly one less.
A counting-to-zero argument can be used to show the correctness of this scheme w.r.t.~blackhole and loop freedom, as $a)$ only the destination may have a distance of zero and $b)$ the source only starts to utilize $F'$ once the path has been established.

\begin{theorem}\label{thm:blackhole}
The reverse update scheme in~\cite{segway} for flows can be made locally verifiable for both blackhole and loop freedom by enhancing it with distance labelings.
\end{theorem}

\section{Removing the Need for Packet Tagging} \label{sec:loopfreedom}

It is sometimes possible to remove the need for packet tagging
(as required by the approach above), and hence also reduce
the number of rules to be stored by the nodes (as they are often
per-tag), by slightly relaxing
the notion of consistency.
Observe that in the previous section, our approach moreover guaranteed so-called per-packet consistency~\cite{DBLP:conf/sigcomm/ReitblattFRSW12}, where a packet will either take the old $F$ or the new $F'$ path, but never a mix of both.
However, such stronger guarantees are not needed in order to guarantee blackhole and loop freedom.

We assume as such that routing is to be performed destination-based along forwarding trees, which in turn have to be blackhole/loop-free.
It was already observed in~\cite{DBLP:journals/tcs/FoersterLSW18} that consistency in this setting can be verified and consistently updated by including the depth of the node $v$ in the forwarding tree in its label.
As such, specifying the parent and the depth suffices.
In a nutshell, a node $v$ waits until its parent $w$ updates, and then only updates if \textsc{depth}$(v)$$=$\textsc{depth}$(w)+1$ is satisfied.
A downside of the above scheme is that it only specifies a single transition from old to new forwarding rules. 
In order for a second and further updates to be performed, the controller needs to again collect acknowledgments that all nodes have switched, inducing unnecessary overhead.
In the previous section and in 2-phase commit schemes in general, one can just create a new tag to avoid such issues, \eg transitioning from $F$ to $F'$ to $F''$ and so on.
Even if $F'$ is never fully implemented, the packet source can transition to $F''$ once its path is fully provisioned.
It seems at first as if the trick of adding increasing version numbers cannot be directly applied to forwarding trees.
In network flows, there is a single node (the source) from which the traffic along the new path originates, whereas in forwarding trees, all nodes can act as sources, potentially sending across combinations of different forwarding trees (in~\cite{DBLP:journals/tcs/FoersterLSW18}: just 2 trees).
However, instead of waiting for the last update to be completed, we can actually mix different subsequent updates, as long as in each intermediate possible time-step the forwarding is performed along a forwarding tree.\footnote{Note that loop freedom is a structural property of the forwarding graph.}
As such, we add version numbers to each label and observe that we only need to obey a larger-than relationship: as long as any of $v$'s neighbors $w$ is a parent in some larger version number $x$, $v$ may switch to its label (tree) with version $x$ if \textsc{depth}$_x$$(v)$$=$\textsc{depth}$_x$$(w)+1$.
Observe that a node can also skip intermediate labels.
Correctness is guaranteed by the invariant that a node will never switch to a smaller version number.\footnote{For practical purposes, an appropriate circular ordering could be defined.}
Nodes using the largest version number form a correct forwarding tree, as they will not forward to nodes in other trees and in each step reduce the distance to the destination.
Next, observe that for all other forwarding trees (version numbers), the next routing hop will decrease the distance in the label of the parent, respectively switch to a higher version number.
Hence, the packet will reach the destination eventually and loops in the current forwarding state can be locally detected as well:
Assume for the sake of contradiction that the forwarding graph contains some loop with no node ringing an alarm (outputting \textsc{no}).
As every node outputs \textsc{yes}, we can follow the routing loop starting from some node $u$, where in each step, we increase the version number or reduce the distance. 
However, when we reach $u$ again\footnote{As we study a structural property, we assume no updates in the meantime.}, $u$ must either have a smaller depth or a higher version number than itself, a contradiction.

\begin{theorem}\label{thm:dest-loop-many}
By augmenting the update scheme from~\cite{DBLP:journals/tcs/FoersterLSW18} with version numbers, s.t.~a node $v$ may update to larger version numbers $x$, if its respective parent $w$ in $x$ is also in version number $x$ and \textsc{depth}$_x$$(v)$$=$\textsc{depth}$_x$$(w)+1$, we obtain a locally verifiable scheme which preserves blackhole and loop-freedom.
\end{theorem}

\section{Discussion: Speed Up and Fault-Tolerance}
\label{sec:faulttolerance}

\noindent\textbf{Potential speed up gains.} Nguyen \etal\cite{segway} showed in their evaluations that decentralized consistent updates can speed up 
updates by up to 45\% at the median, in realistic scenarios.
We briefly analyze what sort of theoretical speed up is possible in extreme cases, from the viewpoint of message propagation delay, where we assume one hop to take unit time.

\begin{figure}[tbp]
	\centering
		\includegraphics[width=0.42\textwidth]{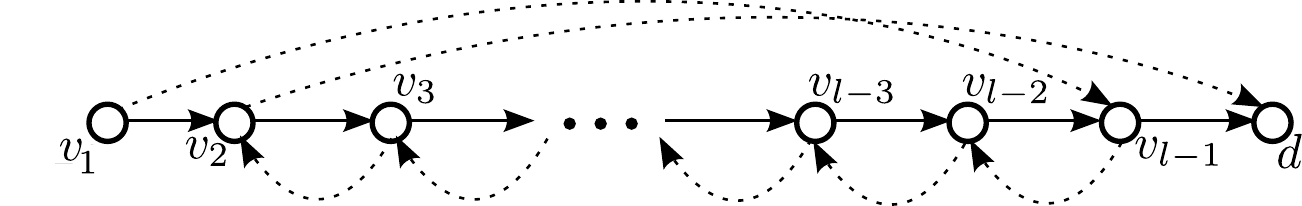}
	\caption{Network with old (solid) and new (forwarding) rules which requires $l-2 \in \Omega(n)$ rounds to update consistently when enforcing loop freedom. For example, $v_3$ cannot update before $v_2$, and so on.}
	\label{fig:WLFvsSLF}
\end{figure}

Consider the scenario analogously to~\cite[Fig.~2]{DBLP:journals/ton/FoersterLMS18}, shown in Figure~\ref{fig:WLFvsSLF}.
The task is to update from the old (solid) to the new (dashed) forwarding rules for the destination $d$ in a loop-free fashion. 
In a centralized setting, we need $\Omega(n)$ rounds to complete the migration, as only one rule (once: two) can be updated per round~\cite{DBLP:journals/ton/FoersterLMS18}.
Else, asynchrony could lead to transient loops in the forwarding graph.
While it is impossible to break the $\Omega(n)$ different updates lower bound, distributed updates can drastically improve the message propagation delay overhead.
Assume that the controller is connected to or placed on any arbitrary node. 
In a distributed setting, the controller can pipeline the distribution of the update labels, reaching all nodes in $O(n)$ time.
Next, the update messages propagate one hop, 
each along the new forwarding rules, again taking $O(n)$ time. 
In contrast, in a centralized setting, the controller needs to obtain an acknowledgement of each update, in turn sending out the next update command. 
In total, this requires a message propagation delay of $\Omega(n^2)$.

\begin{observation}
Distributed updates can speed up the update process by a factor of $O(n)$, w.r.t.~message propagation delay.
\end{observation}
\vspace{-0.2cm}
\noindent\textbf{Fault-tolerance.}
Fault-tolerance is largely unexplored w.r.t.\ proof labeling schemes, the only work that we are aware of relies on a global (unspecified) notification that an error occurred~\cite{DBLP:conf/icdcn/FoersterRSW17}, investigating a single link failure.
On the other hand, there is also work that studies so-called \emph{local fixing}, where nodes/links can \eg leave last wills behind in order to restore properties~\cite{DBLP:conf/opodis/KonigW13}.
However, such fixing is not studied from the aspect of verification, to the best of our knowledge.

Interestingly, we can create a heuristic that directly extends our constructions from the last section to fault-tolerance.
For destination-based routing, observe that we do not need to forward to a node with a depth exactly one smaller, but any smaller depth (or higher version) would suffice. 
In this context, fault-tolerance could benefit benefit from link-disjoint forwarding trees~\cite{DBLP:journals/ton/ChiesaNMGMSS17}, which can be computed efficiently~\cite{DBLP:conf/soda/BhalgatHKP08}, along with appropriate optimization for route lengths~\cite{dsn19,srds19}.

\section{Related Work}
\label{sec:relwork}
Proof labeling schemes have been widely studied in the context of distributed computing. We take inspiration from, \eg\cite{DBLP:journals/dc/KormanKP10,DBLP:journals/toc/GoosS16}, and also refer to both articles for an introduction to the topic.
Similarly, the topic of consistent network updates in SDNs has received much attention in the networking community, see the recent survey in~\cite{survey}.
The idea to leverage proof labeling schemes for verification purposes in SDNs was first investigated in~\cite{DBLP:conf/sigcomm/SchmidS13}, joined with consistent updates for destination-based routing in~\cite{DBLP:journals/tcs/FoersterLSW18}.
We extend the ideas in~\cite{DBLP:journals/tcs/FoersterLSW18} by handling multiple subsequent updates and also covering flow-based routing, along with speed ups and fault-tolerance.
%
%
%

%
Nguyen \etal\cite{segway} lay the practical groundwork for our paper, by showing how to efficiently implement consistent SDN updates in the data plane.
We build upon their work by adding local verification to blackhole and loop-free consistent updates, leveraging the concepts of proof labeling schemes.

Lastly, the idea of fault-tolerance in proof labeling schemes was considered in~\cite{DBLP:conf/icdcn/FoersterRSW17}, but in contrast required an explicit (unspecified) global failure notification.
Related in this context is also the idea of local fixing~\cite{DBLP:conf/opodis/KonigW13} or preprocessing in distributed control planes in general~\cite{DBLP:conf/sigcomm/SchmidS13,supported-local,supported-congest}.

\section{Conclusion}
\label{sec:conclusion}

Given the constantly changing demands and requirements
on communication networks, e.g., due to security policy changes,
traffic engineering requirements, planned maintenance work
or unplanned link failures, among many more,
future communication networks are expected to be 
changed and reconfigured more frequently.
This paper presented a distributed approach,
based on proof labeling systems, which allows to 
offload the responsibility for network reconfigurations
to the data plane and hence support and speed up 
such reconfigurations.

We understand our work as a first step, and believe
that it opens several interesting avenues
for future research.
In particular, it will also be interesting to consider the use of 
randomized~\cite{DBLP:journals/dc/FraigniaudPP19} and approximate~\cite{CENSORHILLEL2018} solutions to improve our approach, provide extensions to further consistency properties such as waypoints~\cite{DBLP:journals/ton/LudwigDRS18} and congestion~\cite{DBLP:conf/infocom/BrandtFW16}, as well as seamless updates~\cite{DBLP:journals/cn/DelaetDKT18}, but also the inherent connections to self-stabilization~\cite{DBLP:journals/tcs/DolevT09}.
More generally, we believe that our approach can provide
interesting new perspectives on emerging 
self-driving networks~\cite{feamster2017and}, which 
center around fine-grained and fast adaptions of networks
reacting to their environment, and may hence benefit
from our distributed approaches.
Furthermore, it will be interesting to investigate
opportunities coming from emerging programmable dataplanes,
to speed up our approach further, as well
as to generalize it to additional use cases.

\bibliographystyle{IEEEtran}
\bibliography{nca19}

\end{document}